\documentclass[twocolumn]{aastex701}

\usepackage{newtxtext,newtxmath}
\usepackage[T1]{fontenc}

\usepackage{amstext}
\usepackage{url}
\usepackage[]{times,refname,amsmath,bm}

\bibpunct{(}{)}{;}{a}{}{,}
\usepackage{tabularx}
\usepackage{graphicx,epsfig,color,latexsym}
\usepackage{orcidlink} 
\renewcommand{\d}{\mathrm{d}}
\newcommand{\e}{\mathrm{e}}

\newcommand{\bea}{\begin{eqnarray}}
\newcommand{\eea}{\end{eqnarray}}
\newcommand{\be}{\begin{equation}}
\newcommand{\ee}{\end{equation}}
\newcommand{\rund}[1]{\left(#1\right)}

\newcommand{\eck}[1]{\left[ #1 \right]}

\begin{document}

\title[Magnification of plasma lensing]{The influence of plasma lensing magnification to the luminosity function of fast radio bursts}
\author[0000-0002-8700-3671]{Xinzhong Er}
\affiliation{Tianjin Astrophysics Center, Tianjin Normal University, Tianjin, 300387, China, phioen@163.com}
\email{}
\author[]{Weisha Zhu}
\affiliation{School of Physics and Astronomy, Sun Yat-Sen University, Zhuhai campus, No. 2, Daxue Road Zhuhai, Guangdong, 519082, China}
\affiliation{CSST Science Center for the Guangdong-Hong Kong-Macau Greater Bay Area, Daxue Road 2, 519082, Zhuhai, China}
\email{zhuwshan5@mail.sysu.edu.cn}
\author[]{Shude Mao}
\affiliation{Department of Astronomy, Westlake University, Hangzhou, Zhejiang, 310030, China}
\email{shude.mao@gmail.com}
\author[0000-0001-7931-0607]{Dongzi Li}
\affiliation{Department of Astronomy, Tsinghua University, Beijing 100084, China}
\email{dzli@tsinghua.edu.cn}

\begin{abstract}
Small scale clumps of ionized gas have been suggested by observations in interstellar medium and circumgalactic medium. The propagation of radio signals can be deflected by these plasma clumps, i.e. plasma lensing. One observable consequence is the magnification and demagnification of background sources. These effects distort the observed luminosity function and potentially introduce bias into population studies. 
In this work, we investigate these effects on fast radio bursts using Gaussian plasma clumps distributed across multiple lens planes within a small field of view. The central electron density for each clump is sampled from uniform, log-normal, and Gaussian distributions. Two analytical models are employed to mimic the intrinsic luminosity function.
Our results show that plasma lensing can modify the observed luminosity functions. 
On one hand, our model shows that radio sources may be demagnified below the detection threshold, the strength varies between $\sim1-15\%$ depending on the ionized gas model and the source redshift. 
On the other hand, magnification can produce anomalously bright sources at the high luminosity end. Both effects introduce potential biases in inferred source properties. The lensing strength correlates with the power spectrum of free electron density. However, scattering effect in the host galaxy or in the Milky Way can suppress the plasma lensing effects. 
\end{abstract}
%
\section{Introduction}
Plasma lensing is a phenomenon that the ionized gas in the universe refracts the photons propagating in it. It was initially proposed to explain the Extreme Scattering Event \citep[ESE,][]{ESE0,1987Natur.328..324R,CleggFL1998}. The plasma acts as an optical medium, refracting the electromagnetic rays in analogy with a conventional concave lens \citep[e.g.][]{2016ApJ...817..176T,2017ApJ...842...35C,2019MNRAS.486.2809G}. Despite a different physical origin, plasma lensing shares mathematics and theory with gravitational lensing. Gravitational lensing probes the mass distribution in the universe by measuring the change of background sources, e.g. the image distortions, source brightness, and positions of multiple images \citep[e.g.][]{SEF92,2022iglp.book.....M}. It also provides a unique way to study objects in the distant universe by the magnification effect. Lensing magnification can alter the observed number counts \citep{1996MNRAS.283.1340B,2010MNRAS.406.2352L,2024JCAP...10..057T} and modify the spectra of the background sources through differential magnification \citep[e.g.][]{2025MNRAS.537.1115X}. Thus, the luminosity function of background sources can be used as a tool to constrain the intervening mass distribution \citep{2005PhRvL..95x1302Z}. Similarly, plasma lensing offers a potential probe of ionized gas in the universe. However, because plasma lensing is frequency dependent, observable signatures are confined to low-frequency regimes, thereby limiting the range of detectable tracers.

Fast radio bursts (FRBs) are radio transients with millisecond duration discovered by \citet{2007Sci...318..777L}. Several lines of evidence are emerging that the FRBs have cosmological origin, such as their excess Dispersion Measure (DM) with respect to the Galactic contribution of free electrons, high brightness, and their isotropic distribution in the sky \citep[e.g.][]{2019ARA&A..57..417C,2019A&ARv..27....4P}. Moreover, the extragalactic origin of FRBs is confirmed by the association of dozens of FRBs with host galaxies at cosmological redshifts \citep[e.g.][]{2017ApJ...834L...7T,2019Sci...365..565B}. Different models have been proposed \citep[e.g.][]{2019ARA&A..57..417C,2021SCPMA..6449501X,2023RvMP...95c5005Z}, but the physical origin of FRBs remains unclear. The rapid growth of the detected sample allows us to study the FRBs statistically, e.g. evolution with redshift or the relation between the event rate of FRB and the star formation rate. Moreover, FRBs provide a potential probe for studying the integrated column density of free electrons along the line of sight \citep[e.g.][]{2021MNRAS.505.5356B,2021ApJ...906...49Z}, and thus for constraining the cosmic baryon content \citep{2014ApJ...780L..33M,2022ApJ...928....9L}. It has been shown that stars and interstellar medium in the galaxy or galaxy cluster is not sufficient to account for the cosmic baryon budget \citep{1998ApJ...503..518F}, but in the circumgalactic medium (CGM) and intergalactic medium \citep[IGM, e.g.][]{2010ApJ...714..320A,2016ARA&A..54..313M}. On the other hand, the baryon fraction in IGM as a function of redshift shows correlation with the DM of FRBs, e.g. the Macquart relation \citep{2020Natur.581..391M,2022MNRAS.509.4775J}.

All cosmological studies utilizing FRBs fundamentally rely on a statistically unbiased sample. However, both gravitational and plasma lensing effects can alter the observations of the FRBs, individually and statistically. Gravitational lensing can boost the brightness of the sources, but its probability is moderate. In contrast, the probability of plasma lensing can be different, since the ionized gas in the universe is not confined to massive haloes but permeates IGM and CGM. 

Some observations indicate that cool clouds can form in CGM down to sub-pc scales \citep[e.g.][]{2021MNRAS.506..877Z}. Such numerous, small-scale plasma clumps can efficiently induce lensing effects. Plasma lensing can perturb the dispersion relation, and more importantly, its diverging refraction property induces both magnification and demagnification to background sources \citep{2022MNRAS.510..197E}. Recent studies further indicate that the observed spatial distribution of FRBs may be suppressed by foreground plasma clumps within the Milky Way \citep{2025arXiv250906721S}, highlighting a critical selection bias.

We explore cosmic magnification induced by plasma lensing in this work. At the current stage, our study focuses on a qualitative demonstration of the cosmic plasma lensing effects, as several uncertainties remain in modelling plasma clumps. One of the major unknowns concerns the density distribution of ionized gas, both within dark matter haloes (galaxies or galaxy clusters) and in IGM. Analytical models have been developed based on the cosmic baryon density, incorporating assumptions regarding the composition and ionization state of the universe \citep{2018ApJ...867L..21Z}. 
In this study, we adopt results from numerical simulations to characterize the interaction of radio signal with ionized gas in CGM and halo environments \citep{2025ApJS..277...43M}, modelling individual plasma clumps with a simple Gaussian density profile. Observational evidence, such as pulsar studies, suggests that multiple ionized structures can lie along the line of sight \citep{2006ChJAS...6b.233P}. We therefore adopt a multi-plane plasma lensing to account for this complexity. However, we do not include scattering effects from the ISM of either the host galaxy or the Milky Way, which can cause angular broadening of point sources and potentially suppress small-scale lensing signatures.

In this work, we simply adopt a 2D Gaussian column density profile to model plasma clumps and calculate the (de)magnification effects by cosmic plasma lensing. Different intrinsic luminosity samples are employed: a broken power-law and a Schechter function. We introduce the basic theory of plasma lensing in Section 2 and present the lensed luminosity function by our toy model in Section 3. The modification to the luminosity function of background sources is discussed in the end.

\section{Basics of Plasma Lensing}

The detailed introduction of plasma lensing can be found in \citet{2016ApJ...817..176T,2017ApJ...842...35C}. Here we outline the basic formalism. Unlike gravitational lensing, where the mass is mainly concentrated in the haloes, the ionized gas is distributed all the way along the line of sight. However, a full ray-tracing approach will consume a large amount of computational time. Thus, we will adopt the thin-lens approximation with multiple planes for plasma lensing. We use $\beta$ and $\theta$ for the angular position on the source plane or on the image plane respectively. The lens equation can be written as 
\be
\beta = \theta - \alpha(\theta) =\theta - \nabla_\theta \psi(\theta),
\label{eq:lenseq}
\ee
where $\psi$ is the effective lens potential and $\nabla_\theta$ is the gradient on the lens plane. In plasma lensing, the lens potential is proportional to the column electron density $N_e(\theta)$
\be
\psi(\theta) \equiv \frac{D_{ds}}{D_d D_s} \frac{\lambda^2 r_e}{2\pi} N_e(\theta),
\ee
where $D_{ds}$, $D_d$, and $D_s$ are the angular diameter distance from the lens to source, from the lens to observer, and from the observer to source, respectively. $r_e$ is the classical electron radius, and the wavelength of the signal is $\lambda$. In reality, the observed wavelength is redshifted due to the expansion of the universe, i.e. $\lambda_{\rm obs} = (1 + z) \lambda_{\rm rest}$, where $\lambda_{\rm rest}$ is the rest frame wavelength. We can adopt two approaches to calculate the cosmic plasma lensing effects. In the first one, we focus on the observational frequency, and in the second one, we start from the initial frequency of emission. Then we calculate the local frequency in each lens plane and the lensing effects at the corresponding redshifts. 
We adopt the Macquart relation \citep{2020Natur.581..391M,2022MNRAS.509.4775J} to mimic the total plasma density and extrapolate to a higher redshift. Here we only consider the contribution from $DM_{\rm IGM}$, which accounts for the free electron residing between galaxies in the universe \citep[e.g.][]{2003ApJ...598L..79I,2004MNRAS.348..999I}. The multiple lens planes will be constructed in the calculation \citep{2022MNRAS.509.5872E,2022MNRAS.515.6198S}. We project the electron density into the redshift bins. 
In this work, we mainly focus on the lensing magnification effect, which can be calculated from the Jacobian $A$ of the lens equation (\ref{eq:lenseq}), i.e. $\mu^{-1}=$det$(A)$. For an axisymmetric lens, it can be simplified as
\be
\mu^{-1}=\frac{\beta}{\theta}\frac{d \beta}{d \theta}.
\ee

\begin{figure}
\centering
\includegraphics[width=8cm]{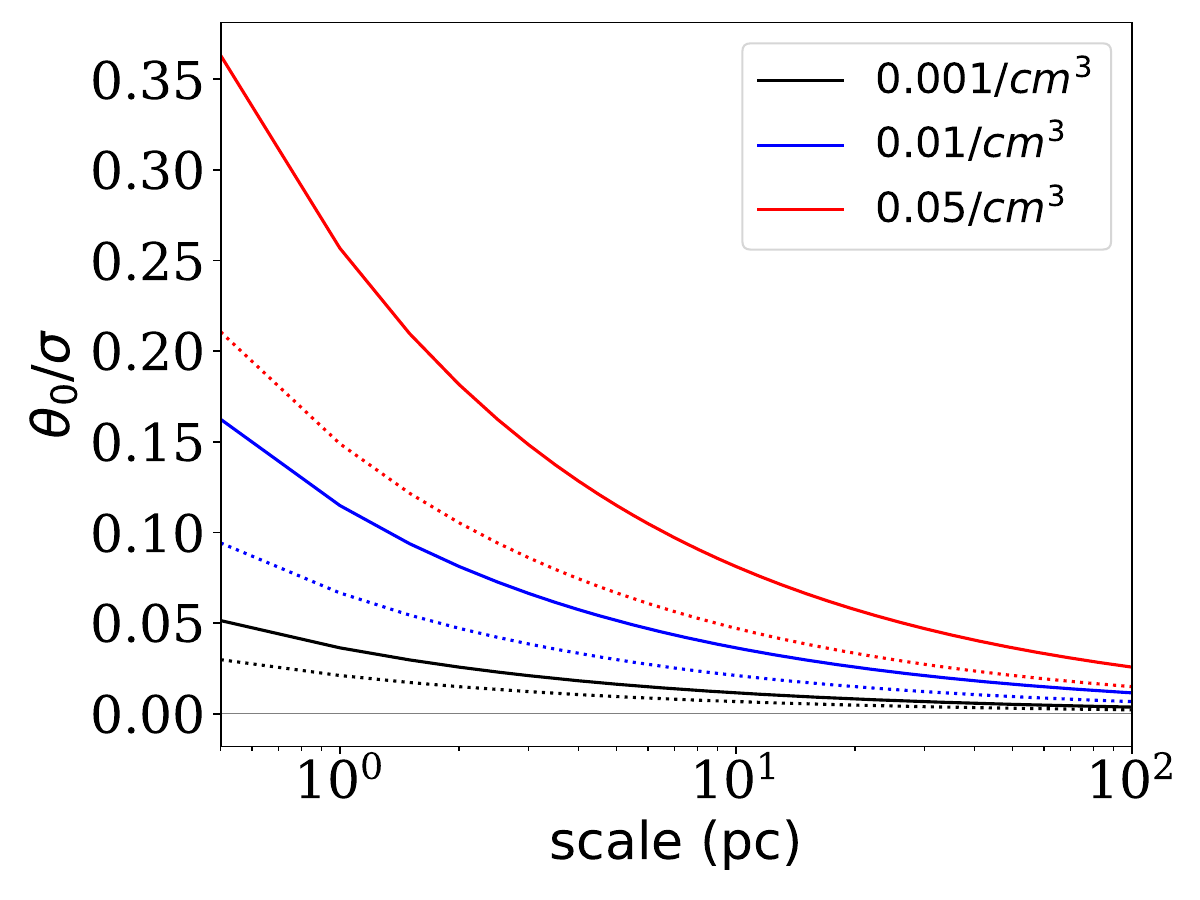}
\caption{The ratio $\theta_0/\sigma$ of plasma lensing as a function of the lensing scale for different electron densities. The solid curves represent the case for a source at $z_s=3.0$ and a lens at $z_d=0.5$, while the dotted curves correspond to a source at $z_s=0.5$ and a lens at $z_d=0.1$. The black, blue and red curves indicate results for the density ($n_e$) of $0.001$cm$^{-3}$, $0.01$cm$^{-3}$ and $0.05$cm$^{-3}$ respectively. }
\label{fig:critical-scale}
\end{figure}

In this work, we will adopt an analytical toy model for the density profile: the widely used Gaussian profile \citep[e.g.][]{CleggFL1998}. It also provides a good description of the Kolmogorov spectrum of the electron density \citep[e.g.][]{2016ApJ...817...16C}. For the axisymmetric Gaussian model, the column electron density of a lens can be written as 
\be
N_e(\theta) =N_0 {\rm exp} \rund{-\frac{\theta^2}{2\sigma^2}},
\label{eq:gauss}
\ee
where $\sigma$ is the width of the lens, and $N_0$ is the maximum column density at the centre of the lens. We define an angular scale $\theta_0$ of the Gaussian plasma lens,
\be
\theta_0^2 = \lambda^2 \frac{D_{ds}}{D_sD_d}\frac{r_e}{2\pi} N_0.
\ee
This angular scale is analogous to the angular Einstein radius in gravitational lensing, but characterizes the ability of a plasma clump to diverge light rays. For a Gaussian clump, the maximum demagnification scales as $\propto(1+\theta_0^2)^{-2}$. The lensing cross section, i.e., the region over which the lens produces demagnification, is determined by the density gradient of the plasma clump. In this case it can be approximated and characterised by the ratio $\theta_0/\sigma$ \citep{2019MNRAS.488.5651E}. Assuming the average electron density in 3D is $n_e$, and the physical scale of the lens is $r_0$, the central column density scales as $\propto n_er_0$, and $\theta_0$ is proportional to $\sqrt{n_er_0}$. Consequently, the ratio $\theta_0/\sigma\propto \sqrt{n_e/r_0}$, indicating that the lensing efficiency decreases with increasing lens size. Other density models, such as power-law profiles, exhibit different dependencies, but follow a similar trend. Plasma lensing starts to produce demagnification (e.g. $\mu<0.95$), when the plasma clump exhibits even a mild concentration, specifically when $\theta_0/\sigma\gtrapprox 0.2$. An example illustrating this condition is shown in Fig.\,\ref{fig:critical-scale}. We can see that for a source located at $z_s=3.0$, a clump with an electron density of $0.05$cm$^{-3}$ and size of $\sim2$ pc satisfies this criterion and can generate significant lensing effects. One needs to notice that this lensing efficiency is for an isolated plasma clump. In reality, with complex larger scale structures, e.g. small clumps within a larger one, the lensing efficiency can be higher than that in Fig.\,\ref{fig:critical-scale}. 


\section{Magnification of plasma lensing}
We employ two approaches to estimate the lensing magnification effects. First, a single Gaussian lens model is used to provide an order-of-magnitude estimate. Second, we perform a simple multi-plane lensing simulation by distributing lenses across several redshift planes to calculate (de)magnification. In both methods, the CGM is treated as the primary contributor to plasma lenses. Although the ISM has a higher electron density and can induce stronger lensing effects, its spatial distribution is highly concentrated within galaxies. In contract, the more extended distribution of CGM results in a higher probability of interacting with radio signals. Consequently, by focusing on the CGM, our model provides a conservative estimate of the overall impact of plasma lensing. 

\subsection{A single Gaussian model}
We present a preliminary estimate of cosmic plasma lensing using a single Gaussian density profile. Ionized gas structures span a broad range of spatial scales, from sub-pc clouds to those associated with massive galactic haloes or intergalactic filaments. However, the mean density of free electron in the IGM is remarkably low, e.g. $\sim10^{-5}$cm$^{-3}$. In order to produce measurable lensing effects with such low density, extremely small scale structures are required. The existence of such small-scale features in the IGM remains a topic of debate. Therefore, we focus on plasma located within or in the vicinity of galactic haloes, i.e., the CGM, where the electron density is significantly higher and the structural complexity is enhanced compared to the IGM. Previous theoretical and simulation studies show that numerous cool ($T\sim 10^4 $ K) gas clumps with sizes ranging from $\sim0.1$ pc and $\sim 100$ pc can form and survive in the CGM (e.g. \citealt{2018MNRAS.473.5407M,2018MNRAS.480L.111G,2020MNRAS.492.1841L}). Meanwhile, observations suggest that the CGM exhibits substantial inhomogeneity on sub-kpc scales, with cool clouds ranging from approximately $0.1$ pc to $1000$ pc required in relevant modelling analyses (e.g. \citealt{2021MNRAS.506..877Z,2024MNRAS.530.3827S,2025arXiv250106551A}).

First, we estimate the lensing probability at the scale of dark matter halos, i.e. ionized gas that traces the dark matter halos of galaxies and galaxy clusters on scales of $\sim10$ kpc to Mpc. Following an approach analogous to gravitational lensing \citep[e.g.][]{2010MNRAS.406.2352L}, we consider only the smooth plasma component following the profile of main halo, i.e. the small scale clumps are not included at this stage. The halo mass function is constructed using the Python package {\it hmf} \citep{2013A&C.....3...23M}. We adopt $M_{200}$ as the halo mass and assume that $1\%$ of the halo mass is in the form of ionized cool gas \citep{2019MNRAS.483..971V}. For simplicity, we approximate ionized gas as a mixture of only protons and electrons to estimate the free electron density. Using the central column density, which can reach up to $\sim10^6$ pc\,cm$^{-3}$ for massive halos, we compute the plasma lensing cross section and integrate over the halo population to obtain the total lensing probability. The resulting probability per halo is found to be low, consistent with our earlier conclusion that contribution of plasma lensing on the scale of dark matter halos is negligible.

According to previous studies \citep[e.g.][]{2025ApJS..277...43M}, a source at high redshift (e.g. $z_s>1.0$) will encounter several haloes in the mass range of $10^{10-14}$M$_\odot$ during its propagation. Assuming that $1\%$ of the halo mass is in the form of ionized cool gas \citep{2019MNRAS.483..971V}, this corresponds to an average free electron density of $\sim0.03$ cm$^{-3}$. While such haloes do not induce significant plasma lensing effect as a whole (with $\theta_0/\sigma\sim10^{-4}$), small-scale clumps within them, down to $\sim1$ pc can contribute. For an intermediate mass halo ($10^{10}$ M$_{\odot}$), the central column density can reach $\sim1000$\,pc\,cm$^{-3}$, whereas the average column density over $r_{200}$ remains $\sim100\,$pc\,cm$^{-3}$. Thus, for a background source, we adopt the DM-z relation to calculate the total column density, and assume that $1\%$ of the ionized gas forms clumps of pc scales within the CGM of these haloes. Under this assumption, a source at $z_s=1.0$ will encounter $\sim50$ small scale plasma lenses on average. This number can vary substantially, and depends on several factors, such as filling factor of plasma clumps.

\begin{figure}
\centering
\includegraphics[width=8cm]{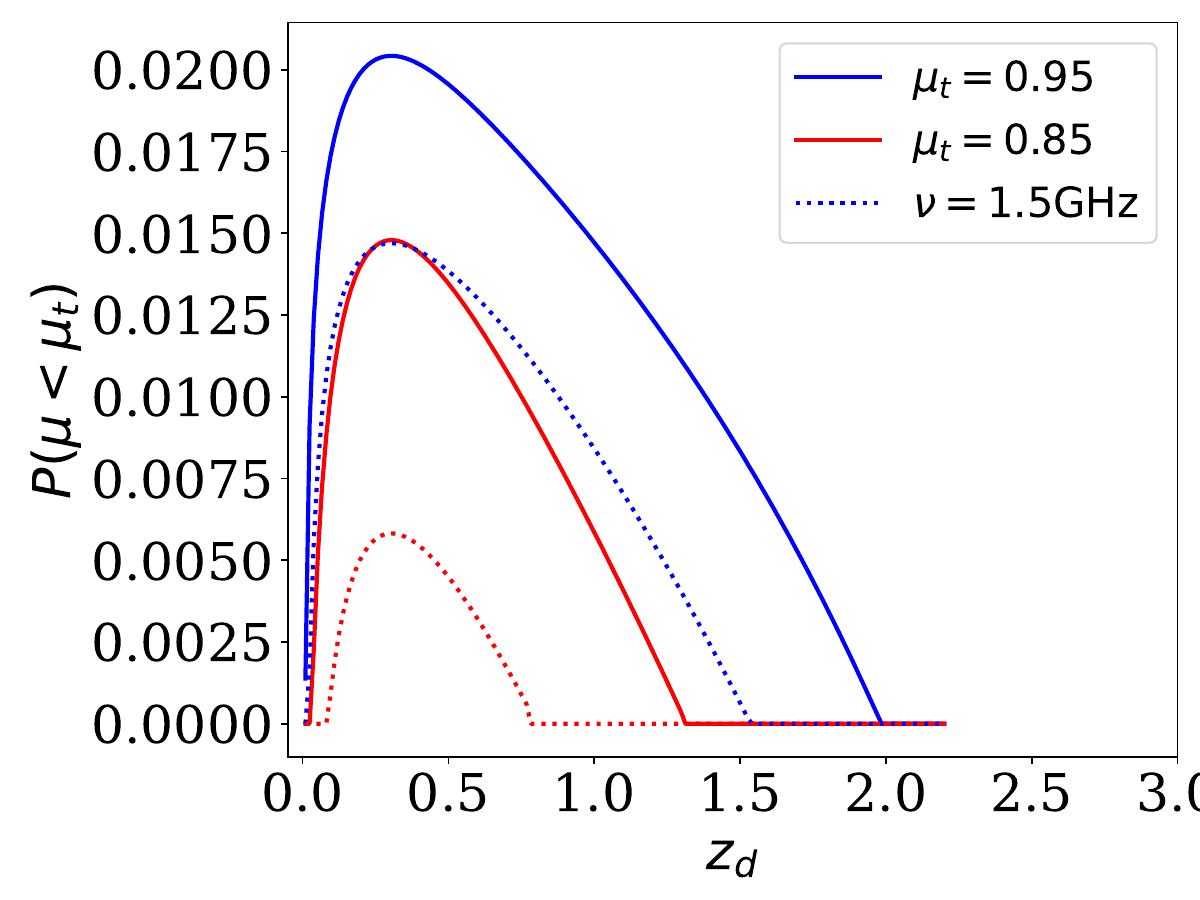}
\caption{The lensing probability for $\mu<\mu_t$ is shown as a function of lens redshift for a single Gaussian profile. The source redshift is $z_s=3.0$. The solid (dotted) curves represent the results for observational frequencies $\nu=1$ and $\nu=1.5$ GHz, respectively.}
\label{fig:single-gauss-prob}
\end{figure}

\begin{figure}
\includegraphics[width=8cm]{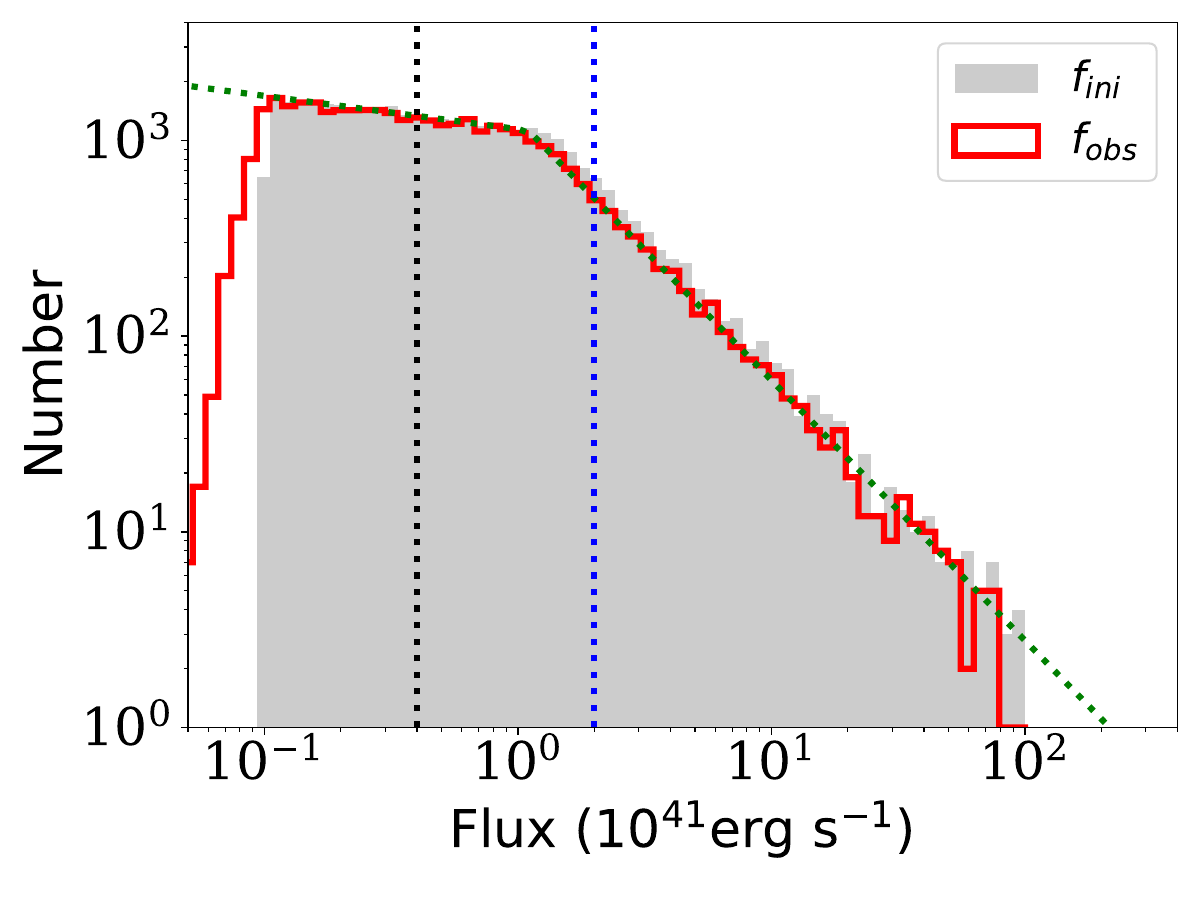}\\
\includegraphics[width=8cm]{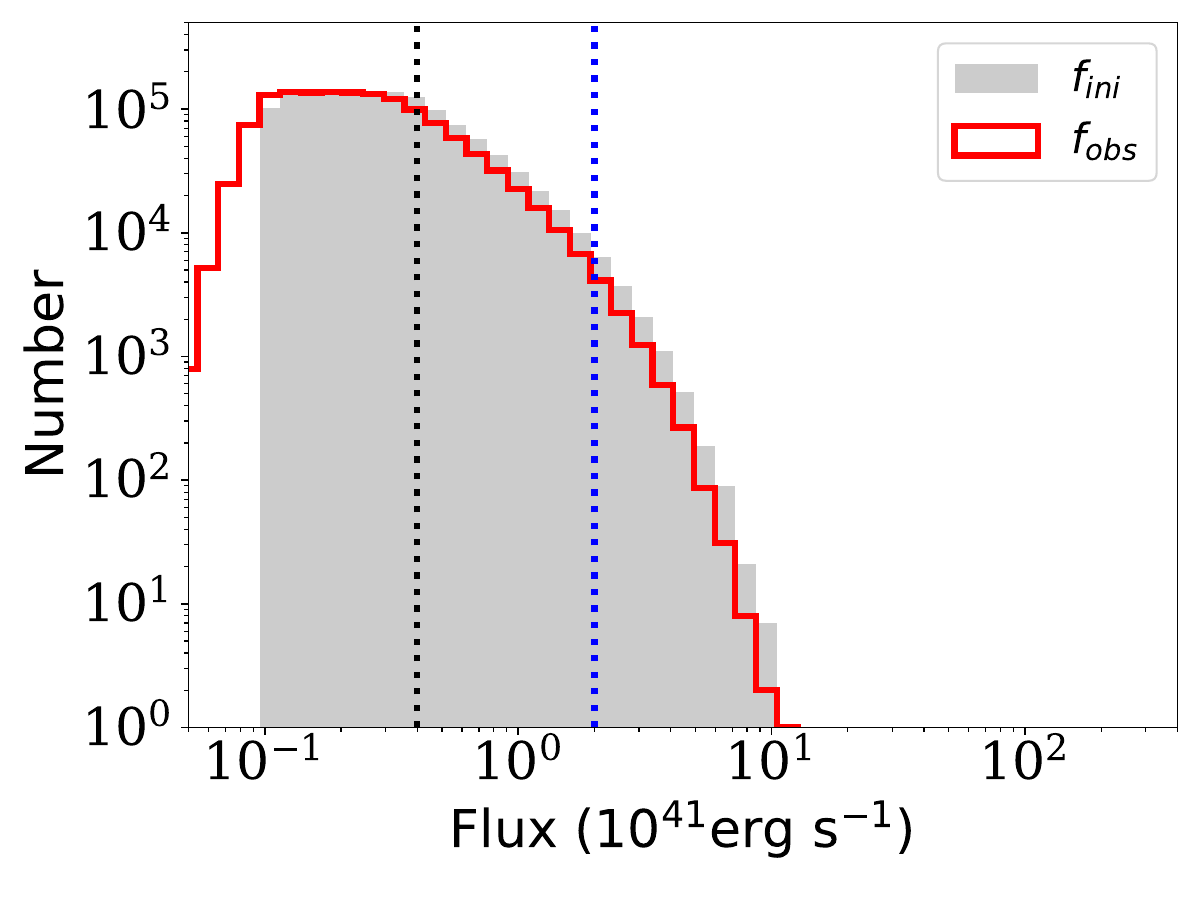}
\caption{The luminosity function before and after a single Gaussian plasma lens. The gray shaded histograms represent the initial source distributions. The red histogram shows the distribution after lensing. The vertical lines mark the mock detection limits. From top to bottom, the initial luminosity functions correspond to the broken power-law and Schechter function, respectively.}
\label{fig:lumz30}
\end{figure}

\begin{figure}
\centering
\includegraphics[width=8cm]{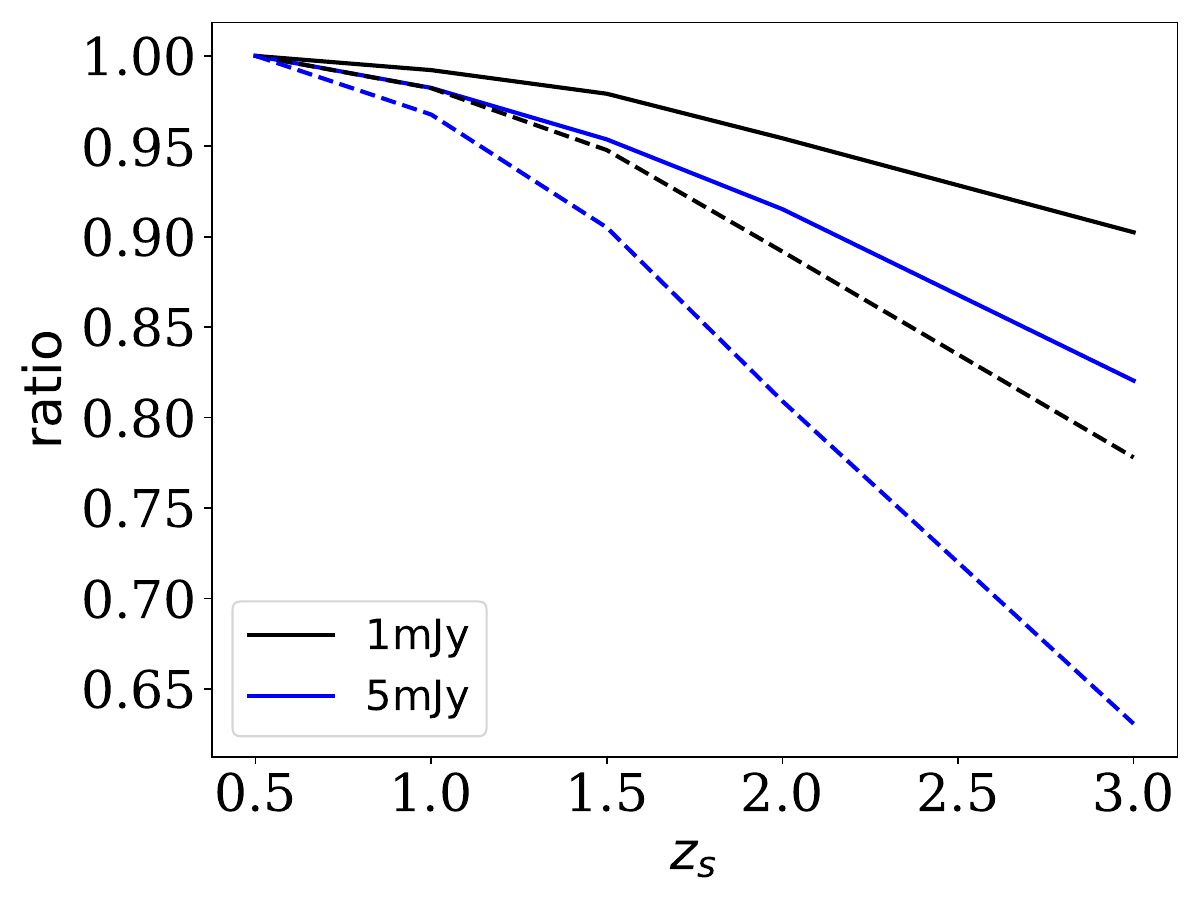}
\caption{The ratio of sources within the detection limit after lensing demagnification. The solid (dashed) lines show the result using a broken-power (Schechter) function. The black (blue) curves show the detection limit of 1 (5) mJy, which corresponding to $0.4$ (2)$\times10^{41}$erg\,s$^{-1}$ at redshift $3.0$. }
\label{fig:gauss-ratio}
\end{figure}

The probability of plasma lensing, both magnification and demagnification, depends on the electron density, size and number of plasma clumps, e.g. $\sim n_e n_{\rm clump}/\sigma$. We compute the lensing cross section for small, Gaussian plasma clumps with $n_e=0.03$ cm$^{-3}$ and $\sigma=1/3$ pc, corresponding to a column density variation of $DM_{\rm clump}\sim0.01$ pc\,cm$^{-3}$. These lenses are subcritical: they cannot produce multiple images, and the cross section for magnification ($\mu>1$) is negligible compared to that for demagnification. In reality, generating multiple images or strong magnification would require higher electron densities in an isolated lens, e.g. $n_e\sim 5$ cm$^{-3}$ \citep{2019MNRAS.488.5651E}.
Thus, in this subsection we focus on the demagnification effect, without multiple imaging. For simplicity, we evaluate the cross section of two demagnification factors $\mu=0.95$ and $\mu=0.85$, and estimate the lensing probability using $r=10\sigma$ as the reference region. Assuming plasma clumps are uniformly distributed along the line of sight, the lensing probability varies with lens distance, peaking when the lens is approximately midway between the source and the observer (Fig.\,\ref{fig:single-gauss-prob}). The dotted curves in Fig.\,\ref{fig:single-gauss-prob} show the results for an observational frequency of $\nu=1.5$ GHz, demonstrating a reduction in strength.

We adopt two analytical models for the background sources. For the first one, we use a broken power-law model \citep{2024arXiv240603672C}, without considering redshift evolution
\be
\psi(L) \propto
\begin{cases}
    (L/L_b)^{-0.17},\quad L<L_b\\
    (L/L_b)^{-1.33},\quad L>L_b,
\end{cases}
\label{eq:lum-fun-ana}
\ee
where $L_b=1.33\times 10^{41}$ erg\,s$^{-1}$ is the break luminosity. We adopt a typical FRB luminosity of order $10^{41}$ erg\,s$^{-1}$, assuming isotropic emission. However, there is suggestion that FRB radiation is highly anisotropic due to beaming effects \citep[e.g.][]{2024ApJ...975..226H}, implying that a large population of sources may intrinsically lie below this luminosity.
To ensure sufficient source counts in each luminosity bin, so that the histogram closely matches the shape of the analytical curve, the simulated sample must contain approximately $35,000$ sources when the lower luminosity limit is set to $0.1\times 10^{41}$ erg\,s$^{-1}$.
Moreover, it has been proposed that the progenitors of FRBs involve neutron stars or stellar size black holes. Thus, the luminosity function of FRBs may be related to the star formation rate and can be described by a Schechter function form \citep[e.g.][]{2023ApJ...944..105S,2025arXiv250109810G}
\be
\psi(L)d L \propto \frac{d L}{L_c} \rund{\frac{L}{L_c}}^{\gamma } {\rm exp} \rund{-\frac{L}{L_c}},
\ee
where $L_c$ is the characteristic luminosity, and we use the same value as $L_b$; $\gamma$ is the power index, and we take $\gamma=-1$. We apply the same lower luminosity limit $0.1\times10^{41}$erg\,$s^{-1}$ to ensure coverage of a broad dynamic range. Because the bright end of the Schechter function declines more steeply than a power law, a relatively large sample is required for the mock distribution to closely trace the analytical shape. Our mock sample contains approximately $1.4\times10^6$ FRBs.

We first investigate background sources at high redshift ($z_s=3.0$). Plasma lenses are modelled as small clumps within the CGM uniformly between the sources and the observer. The total plasma density in these lenses accounts for only a small fraction of the total column density along the line of sight, as inferred from the DM-z relation \citep{2020Natur.581..391M}. 
Although the lensing probability of an individual clump is small (Fig.\,\ref{fig:single-gauss-prob}), the integrated contribution from multiple lenses along the line of sight becomes substantial and complex. For example, in a system with 100 planes of lenses, the cumulative probability for a demagnification of $0.95$ can reach $\sim60\%$. A non-negligible probability is also expected for magnification effect, which will be presented in the following section. This implies a high likelihood that plasma lensing can influence low-frequency flux measurements of high-redshift sources. We present a comparison between the intrinsic and lensed luminosity functions for $z_s=3.0$ in Fig.\,\ref{fig:lumz30}. 

For both samples, plasma lensing shifts the distribution to lower flux values, i.e. it increases the number of low flux sources. In both cases, the demagnification does not change the shape of the function significantly, e.g. it can be represented by the same equation but with slightly modified parameters. In the top panel of Fig.\,\ref{fig:lumz30}, the green dotted line shows Eq.\,\ref{eq:lum-fun-ana} with $L_b=1.05\times10^{41}$ erg\,s$^{-1}$. One reason for this is that we only consider the demagnification effect. In the next part, we will see that the magnification can change the shape of the distribution. We further apply a mock detection limit to calculate the number of sources lost due to lensing demagnification. Two limits are adopted for each sample (vertical lines in Fig.\,\ref{fig:lumz30}). For both samples, we use the same detection threshold of 5 mJy and 1 mJy, corresponding to $2\times10^{41}$ erg\,s$^{-1}$ and $0.4\times10^{41}$ erg\,s$^{-1}$ at $z_s=3$, respectively.
Moreover, we place the sources at different redshifts, from $z_s=1.0$ to $3.0$.  We calculate the ratio of loss due to lensing demagnification, and show its dependence on redshift in Fig.\,\ref{fig:gauss-ratio}. As expected, the loss ratio increases with $z_s$, and higher detection thresholds lead to higher loss ratios. In particular, for the Schechter sample, even sources at $z_s\sim1$ exhibit slight losses.


\subsection{Kolmogorov medium}
For simplicity, we still adopt a Gaussian profile for each plasma clump, and focus on the structures within the foreground haloes and the CGM. Numerical simulations show that radio signals from high-z ($>1$) sources encounter more than 10 haloes during their propagation \citep{2025ApJS..277...43M}. Since we focus on sources at a high redshift, $z_s=3.0$, we place five lens planes uniformly in the redshift interval $[0.01,1.5]$. This range covers the region of peak lensing efficiency near the midpoint between source and observer (Fig.\,\ref{fig:single-gauss-prob}). 

\begin{table*}[]
\centering
\caption{The lens plane and the ratio of sources reduced by lensing de-magnification. The detection limit is 0.2 mJy, which corresponds to about $0.08\times10^{41}$ erg/s at $z_s=3.0$. We use $DM$ to stand for electron column density. $\sigma$ represents the width of the lens (Eq.\,\ref{eq:gauss}). }
\begin{tabular}{c|c|c|c|c|c|c|c}
model &DM dist.  & $\sigma$ (arcsec)  &$\sigma$ dist. & $DM_{\rm lens}$ (pc\,cm$^{-3}$)  &$DM_{\rm lens}/DM_{\rm tot}$  &broken power &Schechter \\
\hline
S1  &log-normal 1 &$10^{-4}-10^{-2}$  &power-law  &2.4  &$0.09\%$   &$1.9\%$  &$2.0\%$  \\
\hline
S2  &Gaussian    &$10^{-4}-10^{-2}$  &power-law  &4.8  &$0.18\%$   &$2.4\%$  &$3.0\%$  \\
\hline
S3  &log-normal 2  &$10^{-4}-5\times10^{-2}$  &power-law  &8.4  &$0.33\%$   &$16.2\%$  &$18.5\%$ \\
\hline
S1u  &log-normal 1  &$10^{-4}-10^{-2}$  &uniform  &3.2 &$0.11\%$  &$4.2\%$  &$5.2\%$
\end{tabular}
\label{tab:ratio}
\end{table*}

We adopt a narrow mass range for the intervening haloes, $10^{10}-10^{12}$ M$_{\odot}$. The plasma structure on each lens plane is modelled as a superposition of two components: a smooth main plasma halo ($>10$ kpc), and a population of small clumps on parsec scales or smaller. We characterize the main halo by its $r_{200}$ with a low electron density of $n_e=0.001$\,cm$^{-3}$ ({e.g. \citealt{2024arXiv241210579C}}). Small clumps are uniformly distributed across the field of the lens plane. The small clump size ($\sigma$) is drawn from either a power law or a uniform distribution. For the small clumps, we adopt $n_e=0.01$ cm$^{-3}$ (e.g. \citealt{2024arXiv241210579C}) to define a mean electron column density, $\langle n_e\sigma\rangle$. The central column density of each small clump is generated from a specific distribution with this mean column density. We employ four different models of the small clumps. The size ranges and other properties of these plasma clump models are summarized in Table\,\ref{tab:ratio}. For some clumps, the central column density reaches $\sim1$ pc\,cm$^{-3}$, yielding $\theta_0/\sigma>0.05$. When combined with the diffuse plasma of the main halo, such clumps can produce non-negligible magnification, though still insufficient to form multiple images.
\begin{itemize}
    \item model S1: angular size ($\sigma$) follows a power-law distribution with power index $\alpha=2$, range $[10^{-4}-10^{-2}]$ arcsec. $10^{-4}$ arcsec corresponds to $\sim0.1$ pc at a few hundred Mpc. The central column density of small clumps follows a log-normal distribution.
    \item model S2: $\sigma$ follows a power-law distribution ($\alpha=2$), range $[10^{-4}-10^{-2}]$ arcsec. The column density follows a Gaussian distribution.
    \item model S3: $\sigma$ follows a power-law distribution ($\alpha=2$), range $[10^{-4}-5\times10^{-2}]$ arcsec. The column density follows a log-normal distribution.
    \item model S1u: $\sigma$ follows a uniform distribution, range $[10^{-4}-10^{-2}]$ arcsec. The column density follows a log-normal distribution.
\end{itemize}

As the lensing magnification is dominated by the small clumps, high spatial resolution is required. Thus, we adopt a lens field of view of $5\times5$ arcsec$^2$, and a mesh of $1024\times 1024$ to perform numerical calculations. $1024$ plasma clumps are randomly sampled in each redshift bin. The mean column density of each lens plane is calculated and summed over five lens planes ($DM_{\rm lens}$ in Table\,\ref{tab:ratio}). For comparison, the total electron column density along the line of sight is estimated from the same $z-DM$ relation used the previous subsection. We adopt a moderate estimate of $DM(z)/z\approx 900$ pc\,cm$^{-3}$ \citep{2004MNRAS.348..999I,2018ApJ...867L..21Z}, yielding $DM_{\rm tot}\approx$2700 pc\,cm$^{-3}$ for a source at $z_s=3$.
The plasma structures responsible for lensing contribute only a small fraction of the total DM, as indicated by the ratio $DM_{\rm lens}/DM_{\rm tot}$ in Table\,\ref{tab:ratio}. We make a conservative assumption about the plasma clumps, since the primary goal of this study is to demonstrate the magnification effect by plasma lensing. Consequently, the mean and variations of our modelled the column density distribution are lower than those in analysis from numerical simulations \citep[e.g.][]{2021MNRAS.505.5356B,2021ApJ...906...95Z}.

\begin{figure}
\centering
\includegraphics[width=8cm]{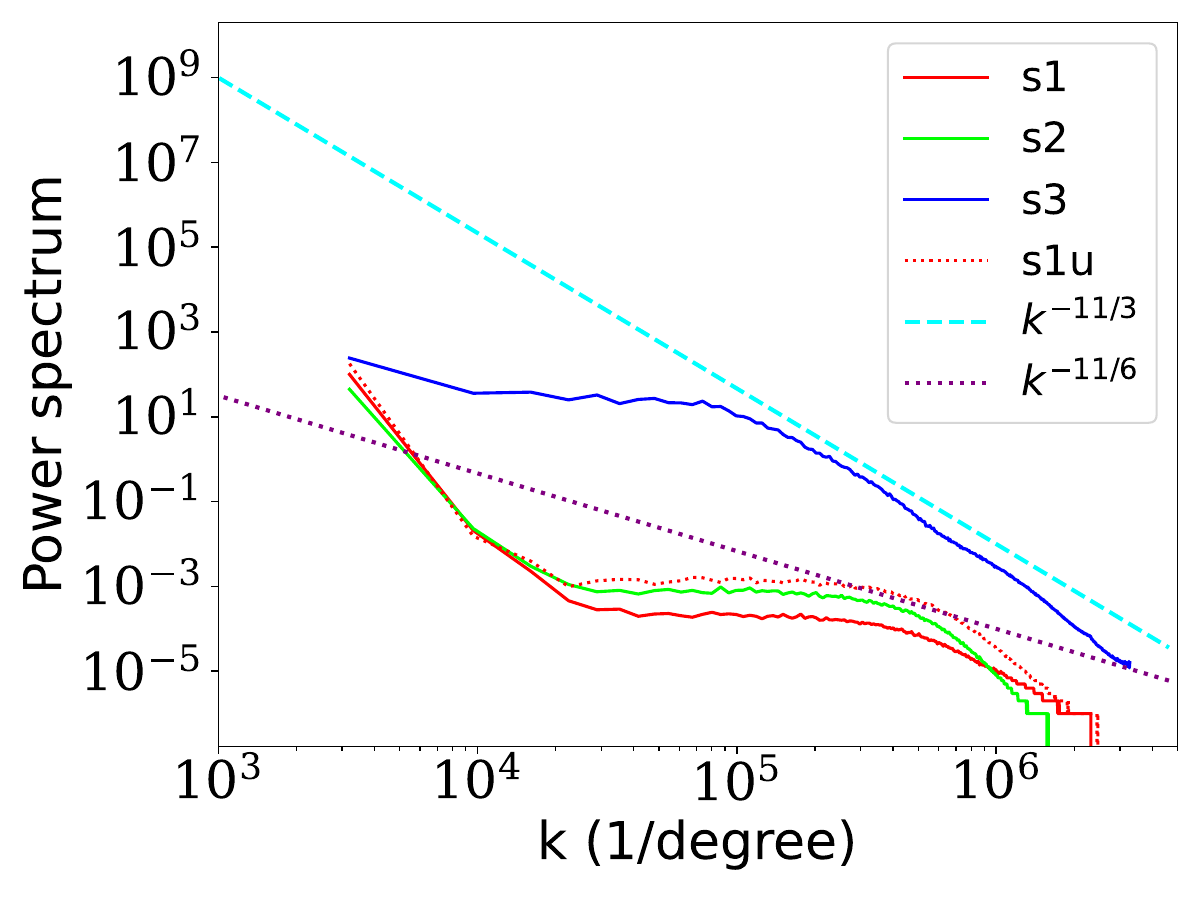}
\caption{The power spectrum of the 2D plasma density in one redshift bin is shown. The purple and cyan lines represent the theoretical power spectrum proportional to $\propto k^{-11/6}$ and $\propto k^{-11/3}$ respectively. $S1-$log-normal, $S2-$Gaussian, $S3-$log-normal with large clumps, $S1u-$log-normal with uniform clump size distribution. }
\label{fig:powers}
\end{figure}

The 2D density power spectra of different distributions are compared with two theoretical models, $\propto k^{-11/3}$ and $\propto k^{-11/6}$. In Fig.\,\ref{fig:powers}, we show an example of the density power of the lenses at $z_d=1.5$. The spectra of samples S1, S2 and S1u share similar slopes, showing power at small scales. S1u has slightly stronger power at intermedia scales than the S1 sample. The S3 sample has a larger size range of plasma clumps, and shows stronger power at small and intermedia scales. The spectrum has a similar slope to the analytical one, $k^{-11/3}$.

\begin{figure*}
\centering
\includegraphics[width=16cm]{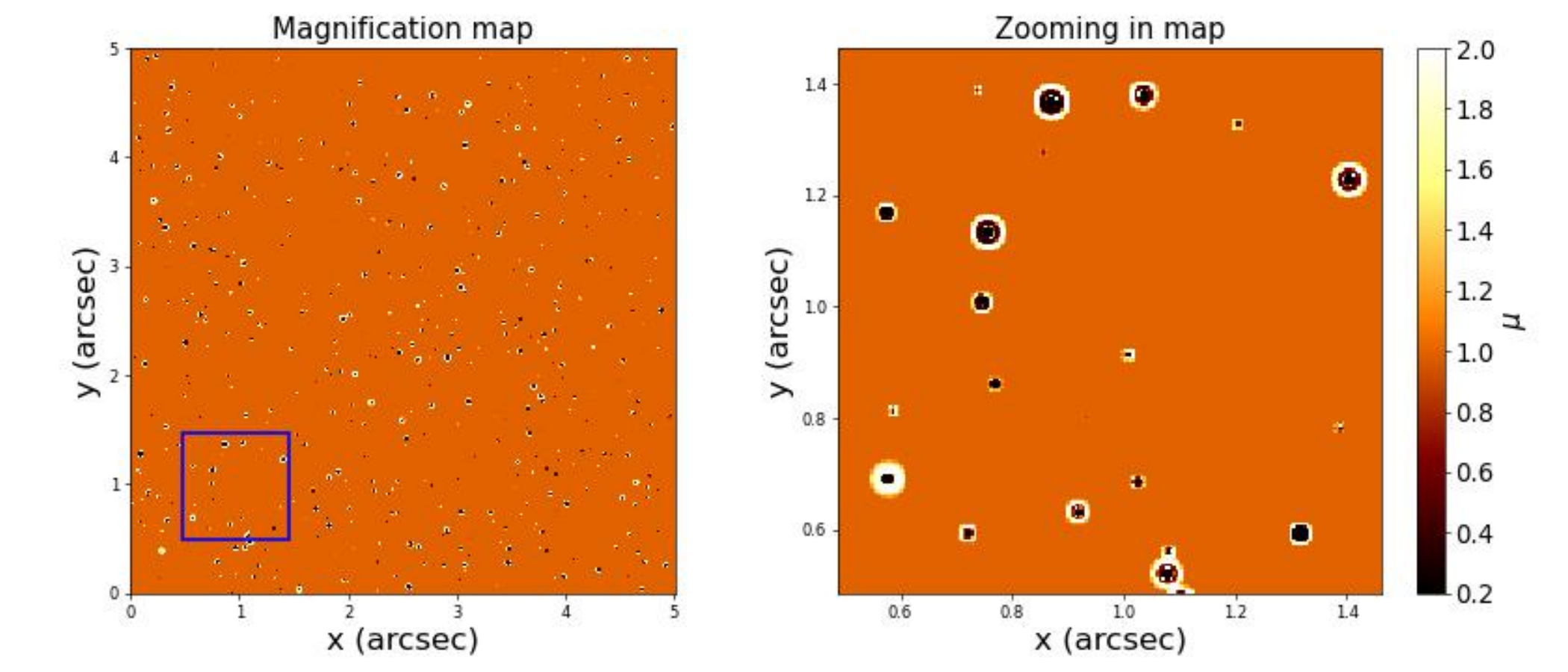}
\caption{Magnification map of one lens plane at $z_d=0.755$ using the S2 sample listed in Table\,\ref{tab:ratio}. The right panel shows a zoom-in of the region indicated by the blue square in the left panel. }
\label{fig:mag-map}
\end{figure*}

Once we have a column density distribution of the plasma in each lens plane, we can calculate the lensing potential. In this section, we focus on the observational frequency of $\nu=1$ GHz; the emission frequency is $\nu(1+z_s)$ from the source, and the frequency at the lens plane of interest in each redshift bin is $\nu(1+z_{di})$. We build the lensing potential accordingly. The lensing magnification is calculated by the numerical differentiation from the potential mesh \citep{2005A&A...437...39B}. We show an example of a magnification map in Fig.\,\ref{fig:mag-map} for a lens plane at $z=0.755$. One can see that the magnifications are small, and do not deviate much from unity. We count the pixels for which the magnification is either greater than $1.5$ or smaller than $0.95$. The ratio for $|\mu|>1.5$ is about $0.2\%$ for models S1, S2 and S1u in one lens plane, and increases to $1.4\%$ for S3. For $|\mu|<0.95$ the ratio is $0.5\%$ for S1, S2 and S1u, and $6\%$ for S3.

The same source samples are used and are uniformly distributed in the field of the lens plane. The magnification in each pixel is used to modify the flux of the source. Since the cross section on the source plane is different from that on the lens plane, we apply a probability of $1/|\mu|$ to the pixel with $|\mu|>1$ to perform the modification of magnification.
The plasma clumps in each redshift bin are not correlated in our simulations. Thus, we do not modify the position of the sources after each lens plane (the deflection angle by plasma lensing is negligible as well). We compare the initial and lensed luminosity functions after all the lens planes.

\begin{figure}
\centering
\includegraphics[width=8cm]{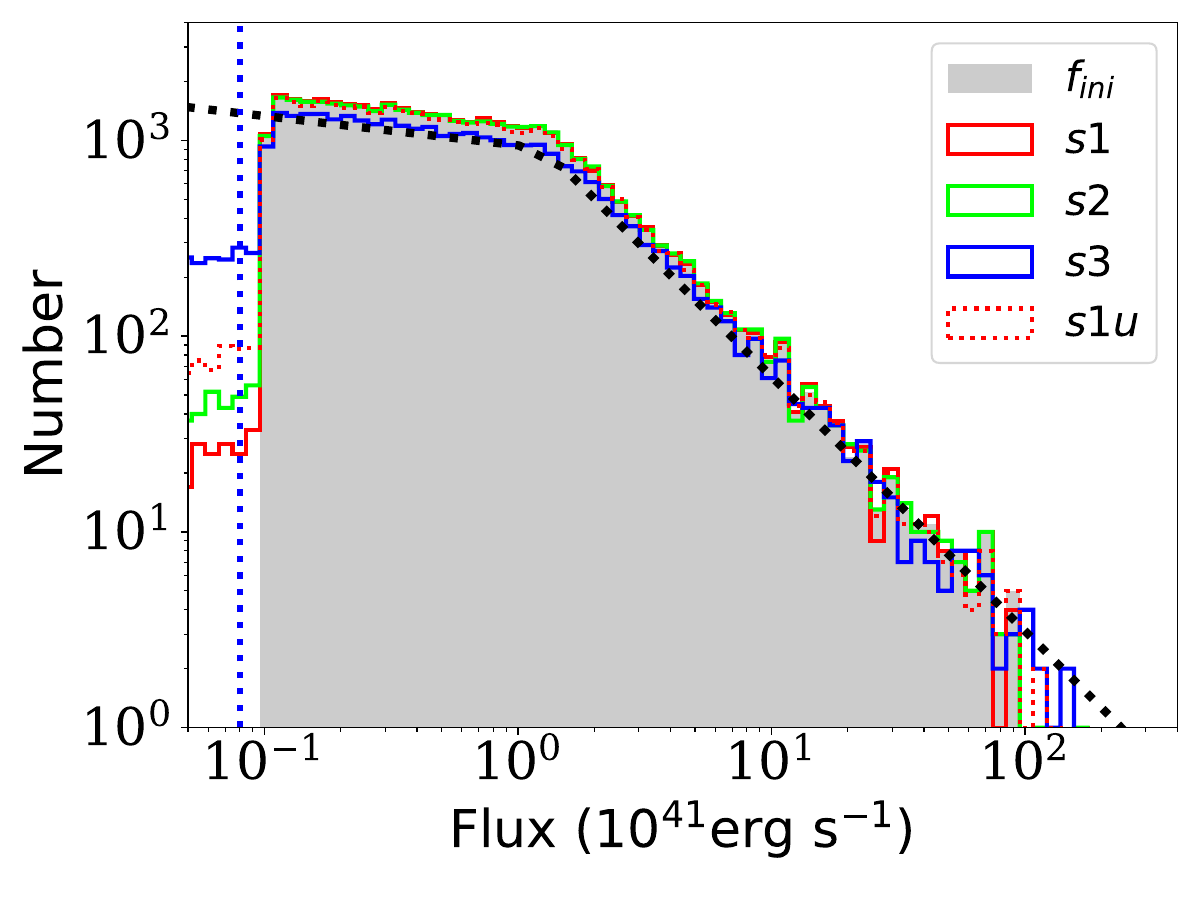}\\
\includegraphics[width=8cm]{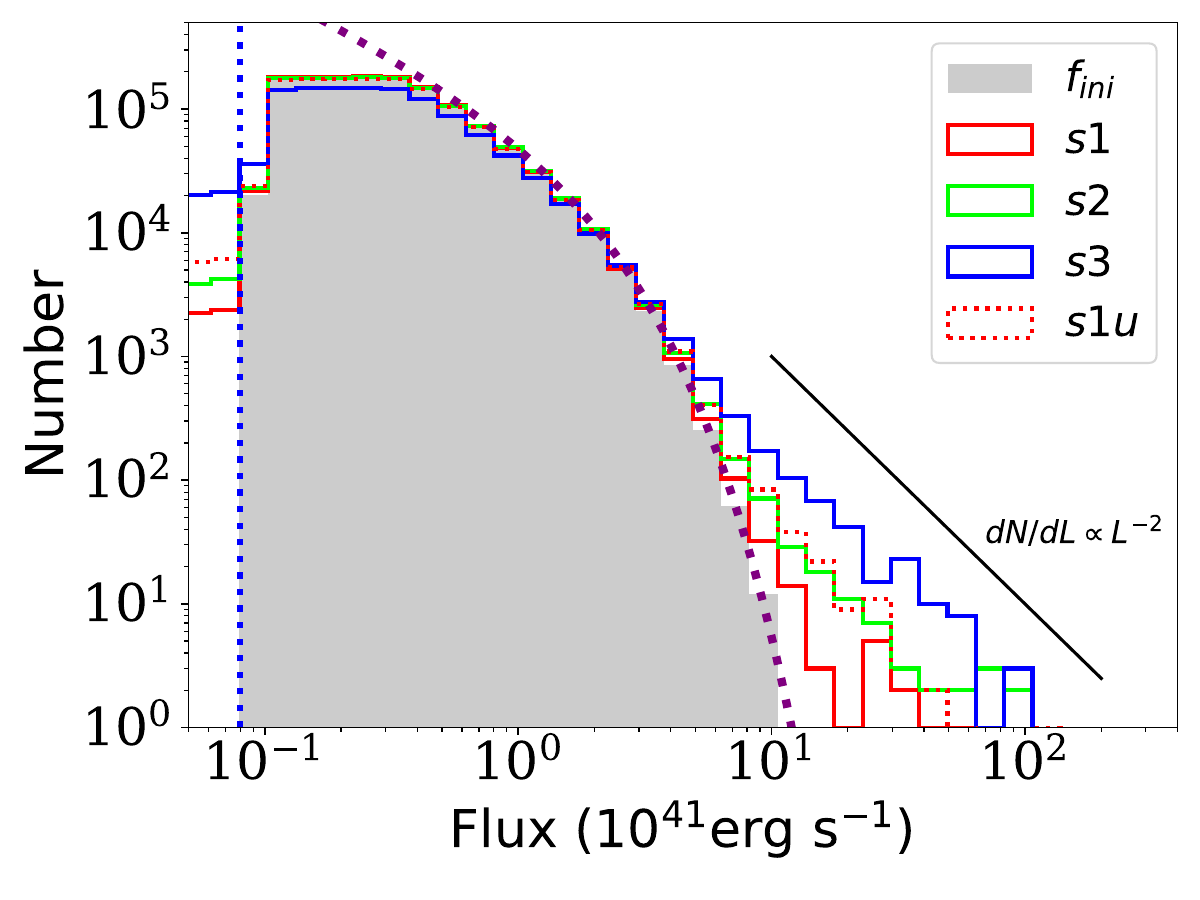}
\caption{Similar to Fig.\,\ref{fig:lumz30}, but using a population of plasma lenses. From top to bottom, the initial luminosity function (grey shaded region) is a broken power-law, and a Schechter function respectively. The labels indicate different models of small plasma clumps. The details are given in Table\,\ref{tab:ratio}. The black line shows an analytical power law relation for comparison.}
\label{fig:kom-lf}
\end{figure}

In Fig.\,\ref{fig:kom-lf}, we compare the luminosity functions with different lensing models. In all tests, the number counts are slightly changed by lensing. First, there are numbers of sources that are demagnified below the minimum flux in the initial sample. The ratio for that is given in the last two columns of Table\,\ref{tab:ratio}. The S1, S2 and S1u lens models slightly reduce the number of sources ($\sim1-5\%$). The S3 model can exceed $15\%$. It is interesting to see that the S1u sample shows slightly stronger lensing effects than S1, as one may expect from the comparison of power spectra. 
In all cases, lensing not only reduces the number of detectable sources, but also changes the slope of the luminosity function slightly.
For example, for the sources following the broken power-law, we show an analytical expression which is almost the same as Eq.\,\ref{eq:lum-fun-ana} by the black dotted line 
\be
\psi'(L) \propto
\begin{cases}
    (L/L_{b'})^{-0.15},\quad L<L_{b'}\\
    (L/L_{b'})^{-1.3},\quad L>L_{b'},
\end{cases}
\label{eq:lum-fun-fit}
\ee
where $L_{b'}=1.3\times 10^{41}$ erg\,s$^{-1}$. The number of sources between $0.1$ and $1\times10^{41}$ erg/s is lower in the S3 sample. The magnification effect boosts the sources slightly toward the bright end of the luminosity function. This effect is particularly significant for a Schechter function, since the intrinsic Schechter function has a relatively steep slope at the bright end. Same as in gravitational lensing \citep{2010MNRAS.406.2352L}, plasma lensing magnification increases the observed abundance of bright sources across all lens models considered.

Although the absolute number of lensed bright sources remains small compared to the initial Schechter sample, their enhanced detectability implies a potentially strong observational bias. Same as power law function, the S3 model produces the strongest lensing effect, causing the bright end slope to approach a power law with an index of $\sim -2$.

Several factors influence the efficiency of plasma lensing. To investigate these effects, we perform additional tests. First, we modify the mean plasma density by introducing an additional homogeneous plasma sheet in each lensing plane. Our results indicate that this alteration has a minimal impact on the (de-)magnification. However, increasing the spatial variation (density gradient) of the plasma density across the field significantly enhances both the spectrum and the lensing effects. Second, the number of plasma clumps exhibits a positive effect on lensing probability, i.e. a large number of clumps leads to a high lensing probability and thus stronger (de-)magnification effects cumulatively. 
Additionally, higher spatial resolution, achieved through a finer computational mesh, results in a small increase in power on small scales and a corresponding small rise in the probability of lensing events.


\section{Discussion and Conclusions}
The content and distribution of ionized gas on various scales in the universe remains poorly understood. This diffuse medium exhibits a substantial difference from both dark matter and stars, which is mainly concentrated within the halo structures. FRBs offer a promising tool for investigating the distribution of the ionized gas in the universe. The observations of a high dispersion measure in numerous FRBs provide compelling evidence for the presence of plasma in the intergalactic medium and circumgalactic medium. Given substantial filling factor in the universe, the probability of plasma lensing is expected to be significant, potentially exceeding that of gravitational lensing. Although there is observational evidence of plasma lensed FRB repeater \citep{2024MNRAS.531.4155C}, estimating the strength of cosmic plasma lensing at current stage is challenging. If we assume that only one plasma clump exists per line of sight (with a comoving area of $\sim$ kpc$^2$), given the small cross section of one plasma lensing, e.g., on the order of $\sim$milli-arcsec$^2$, it reduces the probability to as low as $10^{-6}$. However, this only account a small fraction of the total electron column density along the line of sight. Numerical simulations and observations suggest that for sources at high redshift, e.g., $z\sim 1$, the accumulated electron column density along the line of sight reaches $\sim700$ pc\,cm$^{-3}$ with variations of $\sim100$ pc\,cm$^{-3}$ \citep{2021MNRAS.505.5356B}. Moreover, there is high probability that the radio signals passes through multiple haloes during propagation, and the circumgalactic medium of each halo contains rich small-scale structures \citep[e.g.][]{2025ApJS..277...43M,2021ApJ...906...95Z,2018MNRAS.473.5407M,2018MNRAS.480L.111G,2020MNRAS.492.1841L,2021MNRAS.506..877Z,2024MNRAS.530.3827S}. As we demonstrated in this work, the small scale structures are efficient in lensing effects. Thus, we expect that the probability of cosmic plasma lensing cannot be neglected.

Plasma lensing can induce both magnification and demagnification effects on the radio background sources, leading to complex modifications of their observed luminosity function. 
We investigate the cosmic (de-)magnification effect arising from plasma lensing. A 2D Gaussian plasma clump is adopted for a single lens model. In the first test, we adopt only a single profile lens. In the second one, we use a population of plasma clumps. In all cases, a reduction in the intensity of the background sources due to plasma lensing has been found. The cumulative effect of lensing leads to an increased reduction in background source intensity with redshift. 
This reduction ranges from a few percent to $\sim15\%$ at $z_s=3$ depending on the lensing model and threshold. The magnification effect causes a signature analogous to gravitational lensing: an increase in the number of bright sources and a flattening of the bright end slope of the luminosity function when the intrinsic distribution drops steeply.

We compare the lensing efficiency across different density distributions. Notably, the power spectrum of the log-normal distribution with wider range of lens size exhibits behaviour similar to the analytical form of $\propto k^{-11/3}$. Correspondingly this distribution presents the highest lensing strength. The 2D power spectra can provide a rough indication of lensing efficiency, but reaching a robust conclusion will require more careful studies. 

Our study focuses exclusively on cosmic plasma lensing effects, acknowledging that our simplified plasma density model can not capture the properties of ionized gas in the universe. Nevertheless, even under simple assumptions regarding plasma properties, our simulation reveals that the plasma lensing show probabilities of altering the luminosity function of background radio sources. However, we emphasize that this work is preliminary and subject to various factors that can influence the results, including the mean electron density, the number of multiple lens planes etc. 
\begin{itemize}
\item The plasma lensing is strongly frequency dependent. A first-order approximation of the magnification can be given by $\mu \sim \eck{1-(\nu_0/\nu)^2}^{-2}$, where $\nu_0$ is a reference frequency, e.g. 1 GHz in this work. In additional tests with $\nu=1.5$ GHz, the lensing effects diminish rapidly. In further tests with even higher frequency ($\nu>2$ GHz), the lensing effects become negligible. Consequently, for observations with a wide bandwidth, the plasma lensing effects will be weakened due to averaging across frequencies. However, the overall impact depends critically on the density of the plasma.
\item The number of multiple lens planes can introduce slight variations in the results. However, the most important factor for the final outcome is the presence of plasma clumps along the line of sight. The probability of encountering such clumps increases with redshift. Thus, for high-redshift radio sources, this implies that plasma lensing becomes a non-negligible effect. 
\item 
In reality, the observational data (e.g. the CHIME/FRB catalogue \citep{2021ApJS..257...59C}) is an accumulated sample from all distances. Unfortunately, the redshift distribution of FRBs is not known. The results presented at the current stage are for only one redshift bin. Moreover, there is suggestion that there are multiple populations of FRB repeaters \citep{2021Natur.598..267L}, which will contaminate the signal of plasma lensing as well. 
\item 
In our simulation, magnification is calculated on the lens plane, with a simple correction applied to the source plane. This approach likely underestimates both the probability of magnification and demagnification. In reality, these probabilities depend sensitively on the density profile and morphology of the plasma clumps \citep[e.g.][]{2019MNRAS.488.5651E}, and may differ from the results presented in this work. Moreover, the multiple images produced by lensing are not accounted for in the current analysis. A complete treatment would require full ray-tracing simulations. We leave this analysis to a future work, which would also enable one to employ more realistic models of ionized gas distributions.
\item 
The effects of scattering are not included in this work. Scattering induced by the ISM within the host galaxy or the Milky Way can slightly broaden the angular size of point sources. This effect may reduce or entirely smooth out the lensing effect, particularly for small-scale plasma structures. Previous studies suggest that scattering occurs predominantly within the host galaxy or the Milky Way \citep[e.g.][]{2022ApJ...931...88C}. 

For the lensing effect to remain observable, the scattering timescale $\tau$ must satisfy 
\be
\tau<\frac{1}{c} \frac{d_2^2}{2d_1} \frac{\lambda^2}{x_2^2},
\ee
where $d_2$ is the distance between the source and the CGM lens, $d_1$ is the distance between the source and the scattering screen in the host galaxy, and $x_2$ is size of the CGM lens \citep[e.g.][]{1998ApJ...505..928G}. In typical conditions, this criterion requires extremely small scattering, on the order of $\tau\sim 10^{-3}$ ns, which is challenging to achieve. In our model, a few tens percent plasma clumps may fail to lens sources due to scattering. A rigorous quantification of its impact on our results requires further investigation.
\item 
The ISM in the Milky Way, especially in the disk, is not considered in this work, since the main purpose of this work is to demonstrate the (de-)magnification effect of plasma lensing. For lenses in the Milky Way, a more realistic model can be adopted \citep[][]{2002astro.ph..7156C,2017ApJ...835...29Y}, and direction-dependent results can be obtained.
\end{itemize}

Plasma lensing introduces complexities in interpreting observational radio data, making it a critical factor to consider in radio studies such as FRBs or quasi-stellar objects (QSOs). At the same time, ionized gas plays an important role in the galactic ecosystem. Probing its density distribution and structural properties is essential for understanding galaxy evolution. Plasma lensing provides a unique probe of the spatial distribution and dynamics of ionized gas in the universe. While this study focuses solely on magnification effects, complementary phenomena, such as time delay, dispersion measures, and multi-band observations can offer additional constraints for advancing our understanding of gas distribution as well as cosmic structure.

Moreover, the abundant small structures in galaxies can induce non-negligible magnification of background radio sources. For strongly lensed radio sources, such as QSOs, this effect introduces additional difficulties in lens modelling, while simultaneously offering new insights to investigate the ionized gas in the lens galaxy.

\section*{Acknowledgements}
We thank the referee for her/his constructive comments and suggestions on the manuscript.
W.S.Z. is supported by NSFC grant No. 12173102. S.M. is supported by NSFC grant No. 12133005.

\bibliographystyle{aasjournalv7}
\bibliography{reference} 

\end{document}